\documentclass[aps,prl,nobalancelastpage,superscriptaddress,twocolumn,longbibliography,nobibnotes]{revtex4-2}

\usepackage[utf8]{inputenc}
\usepackage[american,british]{babel}
\usepackage[T1]{fontenc}
\usepackage{graphicx}
\usepackage{xcolor}
\usepackage{dcolumn}
\usepackage{physics}
\usepackage{braket}
\usepackage{bm}
\usepackage{amsmath,amsthm,amssymb}
\usepackage{color}
\usepackage{verbatim}
\usepackage[normalem]{ulem}

\usepackage{hyperref}
\hypersetup{
 colorlinks=true,
 linkcolor=blue,
 anchorcolor = blue,
 citecolor = blue,
 filecolor = blue,
 urlcolor = blue
}

\def \be {\begin{equation}}
\def \ee {\end{equation}}

\newcommand{\iop}{Institute of Physics, Chinese Academy of Sciences, Beijing 100190, China}
\newcommand{\lptms}{Universit\'e Paris-Saclay, CNRS,  Laboratoire de Physique Théorique et Modèles Statistiques, 91405, Orsay Cedex, France.}
\newcommand{\iuf}{Institut Universitaire de France, 75005, Paris, France}
\newcommand{\iogs}{Universit\'e Paris-Saclay, Institut d’Optique Graduate School, CNRS, Laboratoire Charles Fabry, 91127 Palaiseau Cedex, France}

\begin{document}

\title{Observing weakly broken conservation laws in a dipolar Rydberg quantum spin chain}

\date{\today}

\author{Cheng~Chen}
\affiliation{\iop}
\affiliation{\iogs}

\author{Luca~Capizzi}
\affiliation{\lptms}

\author{Alice~March\'e}
\affiliation{\lptms}

\author{Guillaume~Bornet}
\affiliation{\iogs}

\author{Gabriel Emperauger}
\affiliation{\iogs}

\author{Thierry~Lahaye}
\email{thierry.lahaye@institutoptique.fr}
\affiliation{\iogs}

\author{Antoine~Browaeys}
\email{antoine.browaeys@institutoptique.fr}
\affiliation{\iogs}

\author{Maurizio~Fagotti}
\email{maurizio.fagotti@universite-paris-saclay.fr}
\affiliation{\lptms}

\author{Leonardo~Mazza}
\email{leonardo.mazza@universite-paris-saclay.fr}
\affiliation{\lptms}
\affiliation{\iuf}

\begin{abstract} 
Integrable quantum many-body systems host families of extensive conservation laws, some of which are fragile: even infinitesimal perturbations can qualitatively alter their dynamical constraints. Here we show that this fragility leaves a clear experimental fingerprint in a one-dimensional quantum spin chain of as few as 14 Rydberg atoms. Weak integrability breaking from interatomic dipolar couplings is directly detectable within experimentally accessible times in the dynamics of non-local observables. In particular, magnetization fluctuations are highly sensitive to the breaking of fragile conservation laws and exhibit anomalous growth, which we observe experimentally; similar signatures appear in a semilocal string observable. Numerical simulations on substantially longer chains and a simplified classical stochastic model reproduce those features. We establish non-local observables as a sensitive probe of fragile conservation laws in quantum spin chains and Rydberg-atom arrays as a platform to test perturbative descriptions of quantum many-body dynamics with weak integrability breaking.
\end{abstract}

\maketitle

\textbf{Introduction}

The description of the physical world in terms of simplified, integrable models that admit exact solutions is a standard strategy in both few-body and many-body physics~\cite{Arnold1978, Korepin1993, takahashi1999thermodynamics, sutherland2004beautiful}. 
Yet, while integrable models are ubiquitous in theoretical works, real systems are never described by them: perturbations are omnipresent and require appropriate mathematical approaches~\cite{Kato1995, Nayfeh2011, Holmes2013}.
Additionally, integrability is not synonymous with analytical simplicity, as exact solutions are often highly nontrivial and rarely translate into simple or intuitive descriptions.~\cite{Baxter2007Exactly}. 
Integrable systems nonetheless remain a fundamental tool because many realistic models can be viewed as small deformations of an integrable one, and the most robust parts of integrable predictions can survive to realistic situations, where more accurate descriptions necessarily involve sophisticated, nonintegrable models~\cite{Gutzwiller1990, Strogatz2024}. 
This regime, known as \textit{weak integrability breaking}, has emerged as a research field in its own right, stimulated by the groundbreaking Kolmogorov–Arnold–Moser (KAM) theorem for classical systems, which shows that for sufficiently weak perturbations integrable and nonintegrable phenomenology can coexist~\cite{Kolmogorov1954, Moser1962, Arnold1963}.

The notion of integrability in quantum many-body systems is less transparent than in classical mechanics~\cite{Caux_2011}, but it is commonly associated with the presence of an extensive number of compatible conservation laws~\cite{Lusher1976Dynamical,Grabowski1995Structure,fendley2025XYZ}.
A central result on the dynamics of integrable system is that, at late times, a translationally invariant state in an isolated environment becomes locally indistinguishable from a generalized Gibbs ensemble (GGE)~\cite{Jaynes1957Information,Rigol_2007}: a stationary state characterized by an extensive set of generalized temperatures associated with the conservation laws (for reviews, see Refs.~\cite{Vidmar_2016, Essler_2016}). Although this extensive family of conservation laws allows the system to retain a detailed memory of the initial state, in practice distinguishing a GGE from a thermal ensemble is often a quantitative rather than a qualitative issue.

The contrast between integrable and nonintegrable dynamics becomes much sharper when the system is prepared in an inhomogeneous state, where an infinite number of currents can flow simultaneously. Indeed, each conservation law is associated with a continuity equation for a local density and the corresponding current~\footnote{Quasilocal conserved quantities will not be considered in this article but are widely discussed in the theoretical literature.}, generalizing the familiar currents of energy and magnetization, which become long-lived and non-decaying~\cite{Antal1999Transport, Bernard_2016}.
Because of this, integrable models display a richly structured form of ballistic transport, in stark contrast to generic nonintegrable systems where, with only a few conservation laws, Euler-scale dynamics is much more limited~\footnote{Also in conventional  hydrodynamics domain-wall initial conditions can generate ballistic profiles $q(x,t)=q(x/t)$ spreading over $O(t)$ regions, which are known as rarefaction fans.}.
This phenomenology in integrable systems is now understood within the framework of generalized hydrodynamics~\cite{CastroAlvaredo_2016, Bertini_2016}, which treats the emergent quasiparticles and their conserved charges as the basic hydrodynamic degrees of freedom. If one message has emerged from these studies, it is that integrability is not merely a mathematical construct but a physical principle with far-reaching consequences for transport and relaxation in low-dimensional systems~\cite{Kinoshita2006, Langen2015Science, Tang_2018, Schemmer_2019, Malvania2021, Wei2022, Dubois_2024, Schuettelkopf2025}.


In this work, we investigate weak integrability breaking in a real quantum many-body system consisting of a one-dimensional array of Rydberg atoms~\cite{Endres2016, Browaeys2020}. 
The setup is routinely modeled as a quantum spin chain with a Hamiltonian that comprises a nearest-neighbor part which is integrable, and weak longer-range couplings that break many of the conservation laws associated to the former.
Previous theoretical studies of one-dimensional quantum spin chains with weak integrability breaking have shown that the time scales associated with nonintegrable perturbations can be extremely long~\cite{Bertini2015Prethermalization,Durnin_2021, Moller_2021, LopezPiqueres_2021, Bastianello_2021}. In such approaches, weak perturbations are often treated within a kinetic framework, leading to a picture in which an integrable, ballistic regime crosses over only very slowly to conventional, diffusive hydrodynamics~\cite{DeNardis_2018_Diffusion, DeNardis_2019_Diffusion, Bastianello_2020, Hubner_2025,  biagetti2026generalised}. As a result, it is commonly believed that the small-scale setups realized in quantum simulators, typically comprising at most a few dozen spins, can only access the phenomenology of the underlying integrable model, which is often effectively noninteracting: finite-size effects are expected to completely obscure any integrability-breaking process.

We present the first combined experimental and theoretical evidence that the breakdown of  conservation laws induced by perturbations to integrable models can be resolved even in small quantum simulators and within experimentally accessible times. 
We study the quantum dynamics of an initial inhomogeneous domain-wall state, in which half of the effective spins are polarized along the $z$ axis and the other half in the opposite direction. This state is particularly advantageous for several reasons: (i) experimentally, thanks to recent advances~\cite{Browaeys2020}, it can be prepared with high fidelity; (ii) numerically, its time evolution can be simulated with moderate computational cost using state-of-the-art matrix-product-state techniques, even for relatively large systems (for an early study see Ref.~\cite{Gobert_2005});
and (iii) analytically, its dynamics under integrable Hamiltonians has been extensively characterized in recent years~\cite{Antal1999Transport, Ljubotina2017Spin,Dynamics2017Misguich,Stephan2017Return,Collura_2018_Analytic,Dubail_2017_SciPost, Scopa_2023, Eisler_2025_Scipost}. 
We show that magnetization fluctuations, quantified by their variance and by a related string-order operator, are extremely sensitive to the breakdown of  conservation laws induced by dipolar couplings and therefore provide a powerful experimental diagnostic of weak integrability breaking on unexpected short time scales.

\textbf{Results}

\begin{figure}
	\centering
	\includegraphics{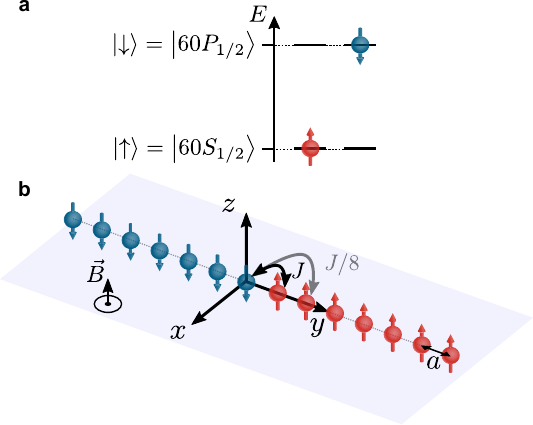}
	\caption{\textbf{Schematic overview of the 1D spin chain.}
	\textbf{a}: Rydberg energy levels of a $^{87}\text{Rb}$ atom used to encode an effective spin-$1/2$. \textbf{b}: Fourteen atoms are arranged in a one-dimensional array with uniform spacing and initialized in a magnetic domain-wall state, defined in the text.} 
	\label{fig:fig1}
        
\end{figure}

Our setup consists of a one-dimensional array of $^{87}\text{Rb}$ atoms trapped in optical tweezers generated by a spatial light modulator~\cite{Nogrette_2014,Schymik_2020}. 
We arrange $L = 14$ atoms in a one-dimensional chain with an equal interatomic spacing $a = 10.8\,\mu\text{m}$, 
and encode a pseudo-spin $1/2$ using two Rydberg states $\ket{\uparrow} = \ket{60S_{1/2},m_J=1/2}$ and  $\ket{\downarrow} = \ket{60P_{1/2},m_J=-1/2}$.
The resonant dipole-dipole interaction allows the system to be described by the following dipolar XX Hamiltonian:
%
\begin{equation}
\label{eq:H}
    \hat{H}_{\rm XX}=-\hbar J
	\sum_{i < j}  \frac{a^3}{r_{ij}^3}  (\hat{\sigma}^+_i \hat{\sigma}^-_j + \hat{\sigma}^-_i \hat{\sigma}^+_j)~,
\end{equation}
with $r_{ij}$ being the distance between site $i$ and $j$ (labeled from $-L/2+1$ to $L/2$), $J= 2 \pi \times 1.2~$MHz the interaction strength, and 
$\hat{\sigma}_i^{x,y}$ the Pauli matrices acting on spin at site $i$. 
To ensure isotropic interactions, we define the quantization axis by applying a $\sim 45~$G magnetic field perpendicular to the array.
The atoms are initialized in a magnetic domain wall state $\ket{\Psi_{\text{DW}}} = \ket{\downarrow\downarrow\downarrow\downarrow\downarrow\downarrow\downarrow\uparrow\uparrow\uparrow\uparrow\uparrow\uparrow\uparrow}$, see Fig.~\ref{fig:fig1}, using the same techniques introduced in our previous works~\cite{Chen2023, Bornet2023,  2025_Chen, Emperauger_2025}; 
more details about the experimental apparatus are presented in \emph{Materials and Methods}.
Starting from this non-equilibrium state, we measure the magnetization $\hat{\sigma}^z_j$ following the evolution governed by the $\text{XX}$ Hamiltonian.
The dynamics of XX models with power-law tails in $d$ dimension has been the object of several recent theoretical and experimental studies, see for instance Refs.~\cite{Frerot_2018, Zunkovic-18, Liu-19, Lerose-20, Kranzl_2023_Noncommuting, Roscilde_2023, Lewis-2023, Defenu-23, Defenu-24, 2025_Chen}.

From the viewpoint of this work, it is crucial to emphasize that the Hamiltonian is dominated by its nearest-neighbor ($r_{ij}=a$) contribution, which coincides with the XX model
\begin{equation}
 \hat H_{\rm nn} = - \hbar J \sum_j
 \left(
 \hat \sigma^{+}_j \hat \sigma^-_{j+1} +
 \hat \sigma^-_j \hat \sigma^+_{j+1}
 \right).
\end{equation}
This model is well known to be integrable: under a Jordan–Wigner transformation it maps to free (noninteracting) fermions, whose quasiparticle excitations each carry one quantum of magnetization~\cite{JordanWigner1928, Sachdev2011}. 
The leading correction to $ \hat H_{\rm nn}$ is the next-to-nearest-neighbor ($r_{i j}=2a$) dipole-dipole term
\begin{equation}
 \hat H_{\rm nnn} = - \frac 18 \hbar J \sum_j
 \left(
 \hat \sigma^{+}_j \hat \sigma^-_{j+2} +
 \hat \sigma^-_j \hat \sigma^+_{j+2}
 \right).
\end{equation}
In the fermionic representation, $\hat H_{\rm nnn}$ generates interactions between quasiparticles and thereby breaks the infinite set of local conservation laws of $\hat H_{\rm nn}$. 
Owing to the rapid $r^{-3}$ decay of dipole-dipole couplings, this platform naturally realizes a controlled regime of weak integrability breaking, as reflected by the small prefactor $1/8$ at $r_{ij}=2a$. Higher-order corrections, such as van der Waals interactions and quadrupolar interactions, also break integrability, but are parametrically smaller and will be neglected hereafter.
A brief discussion of 
quantum integrability and of its relation with the fermionic representations of $\hat H_{\rm nn}$ and $\hat H_{\rm nnn}$ is presented in \emph{Materials and Methods}.

\paragraph{\textbf{Dynamics: Magnetization profile --}}
It is well established that, under the integrable nearest-neighbor Hamiltonian $\hat H_{\rm nn}$, the domain-wall state $\ket{\Psi_{\rm DW}}$  relaxes ballistically: the net magnetization transferred across the junction grows linearly in time~\cite{Antal1999Transport}. This behaviour reflects the free-fermion nature of the XX model: its quasiparticles are stable, propagate with well-defined group velocities, and undergo only elastic (indeed, trivial) scattering. Consistently, the local magnetization develops a light-cone structure, which can be interpreted as the quantum-coherent melting of the initial magnetic order. In the thermodynamic limit and at long times, the magnetization profile develops a scaling form; for $|j|\leq 2Jt$ it behaves as $\langle \hat \sigma^z_j \rangle(t) = (2/\pi)\arcsin(j/(2Jt))$, with saturation to the initial values outside the light cone (see  \emph{Materials and Methods} for more details).

\begin{figure}[t]
\includegraphics[width=\columnwidth]{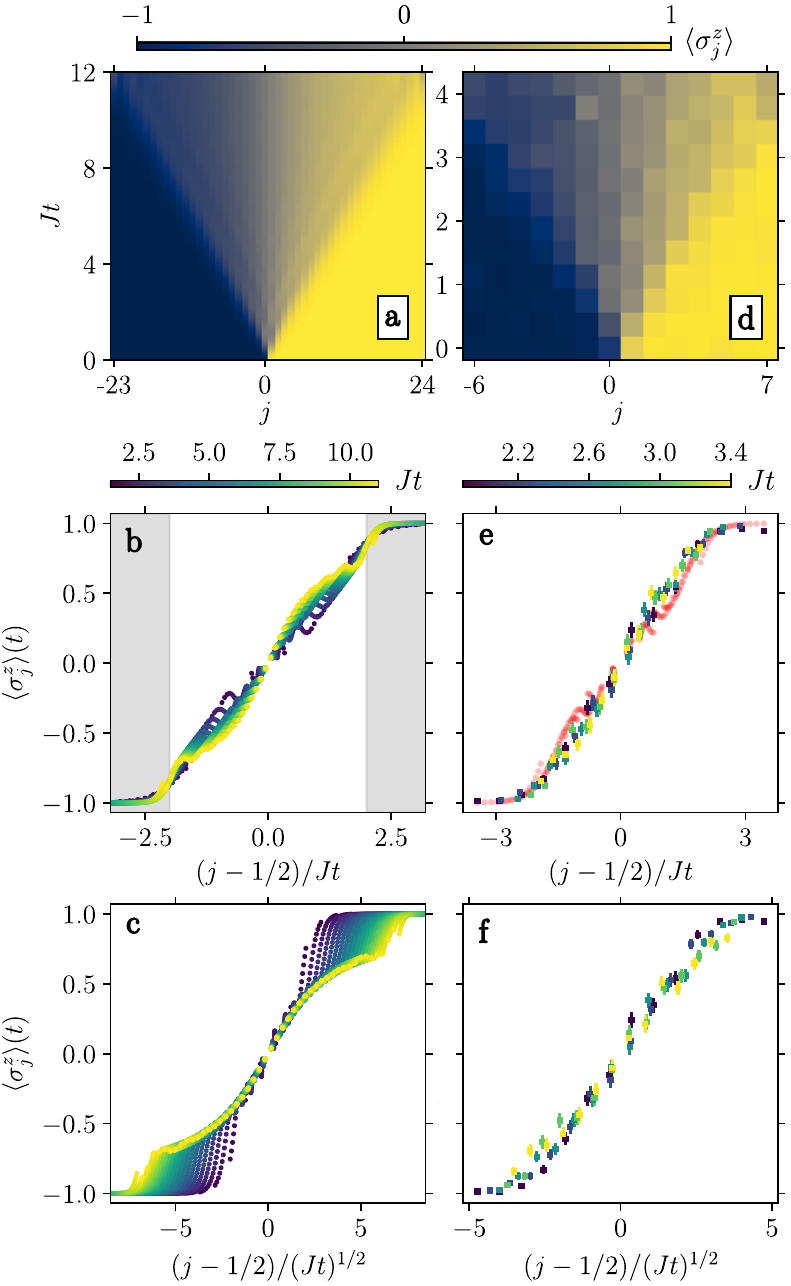}
 \caption{
 \textbf{Magnetization profile.}
\textbf{a}: Space-time map of the numerical magnetization, $\langle \hat{\sigma}^z_j\rangle(t)$, evolved under $\hat H_{\rm nn}+\hat H_{\rm nnn}$ for a chain of length $L=48$. The maximum time is chosen such that finite-size effects remain negligible.
\textbf{b}: Ballistic rescaling of the data in \textbf{a}, plotting $\langle \hat{\sigma}^z_j\rangle(t)$ versus $(j-1/2)/(Jt)$; the gray region indicates the range where the collapse is observed.
\textbf{c}: Diffusive rescaling of the data in \textbf{a}, plotting $\langle \hat{\sigma}^z_j\rangle(t)$ versus $(j-1/2)/(Jt)^{1/2}$.
\textbf{d}: Space-time map of the experimental magnetization, $\langle \hat{\sigma}^z_j\rangle(t)$.
\textbf{e}: Ballistic rescaling of the experimental data; red dots show numerical simulations for $L=14$ evolved under $\hat H_{\rm nn}+\hat H_{\rm nnn}$ over the same time window as the experiment.
\textbf{f}: Diffusive rescaling of the experimental data, plotting $\langle \hat{\sigma}^z_j\rangle(t)$ versus $(j-1/2)/(Jt)^{1/2}$.
}
 \label{Fig:Magnetization}
\end{figure} 

In this setting, the $z$-component of the magnetization $\hat \sigma^z_{\rm tot} = \sum_j \hat \sigma^z_j$ plays a special role: it is an exactly conserved quantity both of the integrable Hamiltonian $\hat H_{\rm nn}$ and of the perturbation $\hat H_{\rm nnn}$.
However, this is very non-generic.
Let us consider another conservation law of $\hat H_{\rm nn}$, such as the magnetization current $\hat J^z_{\rm tot} = i J \sum_j (\sigma^+_j \sigma_{j+1}^- - \text{H.c.}) $:
this is a \textit{fragile}
conservation law of the XX model~\cite{Burgarth2021KAM}, being easily destroyed by generic perturbations and in particular by $\hat H_{\rm nnn}$.
In \emph{Materials and Methods} we recall the important fact that the conservation laws of $\hat H_{\rm nn}$ which are odd under spin flip, such as $\hat J^z_{\rm tot}$, cannot be deformed even at first order under the action of a perturbation ($\hat \sigma^z_{\rm tot}$ is an exception), and are considered fragile.
Without entering into the more involved technical discussion of integrability breaking, we want here to investigate whether the breaking of fragile conservation laws induced by $\hat H_{\rm nnn}$ leaves an observable fingerprint in the experimental data.

We numerically simulate the time evolution of $\ket{\Psi_{\rm DW}}$ under $\hat H_{\rm nn}+\hat H_{\rm nnn}$ in order to quantify how the dipolar next-to-nearest-neighbor coupling modifies the domain-wall melting. We consider a chain of $48$ spins and restrict the evolution to $J t_f = 12$ to minimize finite-size effects. 
Numerics are performed with a standard algorithm based on matrix-product-states (MPS)~\cite{Schollwock2011} implemented in the Julia-based ITensors package~\cite{itensor,itensor-r0.3}, see \emph{Materials and Methods} for more details.
The resulting magnetization profiles $\langle \hat \sigma^z_j\rangle(t)$ are shown in Fig.~\ref{Fig:Magnetization}\textbf{a}. A clear causal light cone propagates from the junction, defined here as the bond connecting sites $j=0$ and $j=1$.
To characterize the transport, we rescale the data using the variable $(j-\tfrac12)/(Jt)^{1/z}$, where $z$ is the dynamical exponent.
Figures~\ref{Fig:Magnetization}\textbf{b} and~\ref{Fig:Magnetization}\textbf{c} display the collapse obtained for $z=1$ (ballistic) and $z=2$ (diffusive), respectively.
We find that the ballistic rescaling yields a nontrivial structure primarily near the light-cone edges, whereas the data show a sharpening increase of the slope around the center ($j\simeq \tfrac12$).
This suggests a tendency towards an almost piecewise-constant plateau, with the development of a discontinuity that is resolved only upon adopting a diffusive scaling. 
We attribute this emergent diffusive broadening to the term $\hat H_{\rm nnn}$, which has broken fragile conservation laws.
In this paper, we will not focus on the resulting two-scale structure, ballistic propagation of the front together with diffusive smoothing near the junction, although it merits further theoretical investigation.

\begin{figure}[b]
\includegraphics[width=\columnwidth]{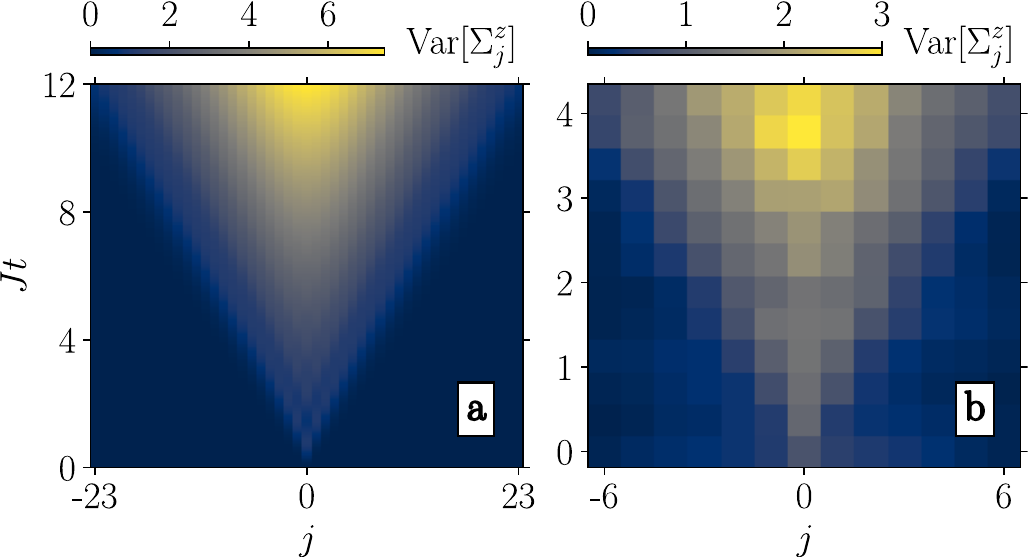}
 \caption{
 \textbf{Variance of the subsystem magnetization: space--time profile.}
 \textbf{a}: Numerical space-time map of $\text{Var}[\hat \Sigma^z_j](t)$ for evolution under $\hat H_{\rm nn}+\hat H_{\rm nnn}$.
 \textbf{b}: Experimental space-time map of $\text{Var}[\hat \Sigma^z_j](t)$ after post-processing (see text).
 }
 \label{Fig:VarianceIntegrMag}
\end{figure}

\begin{figure*}[t]
\centering
\includegraphics[width=\textwidth]{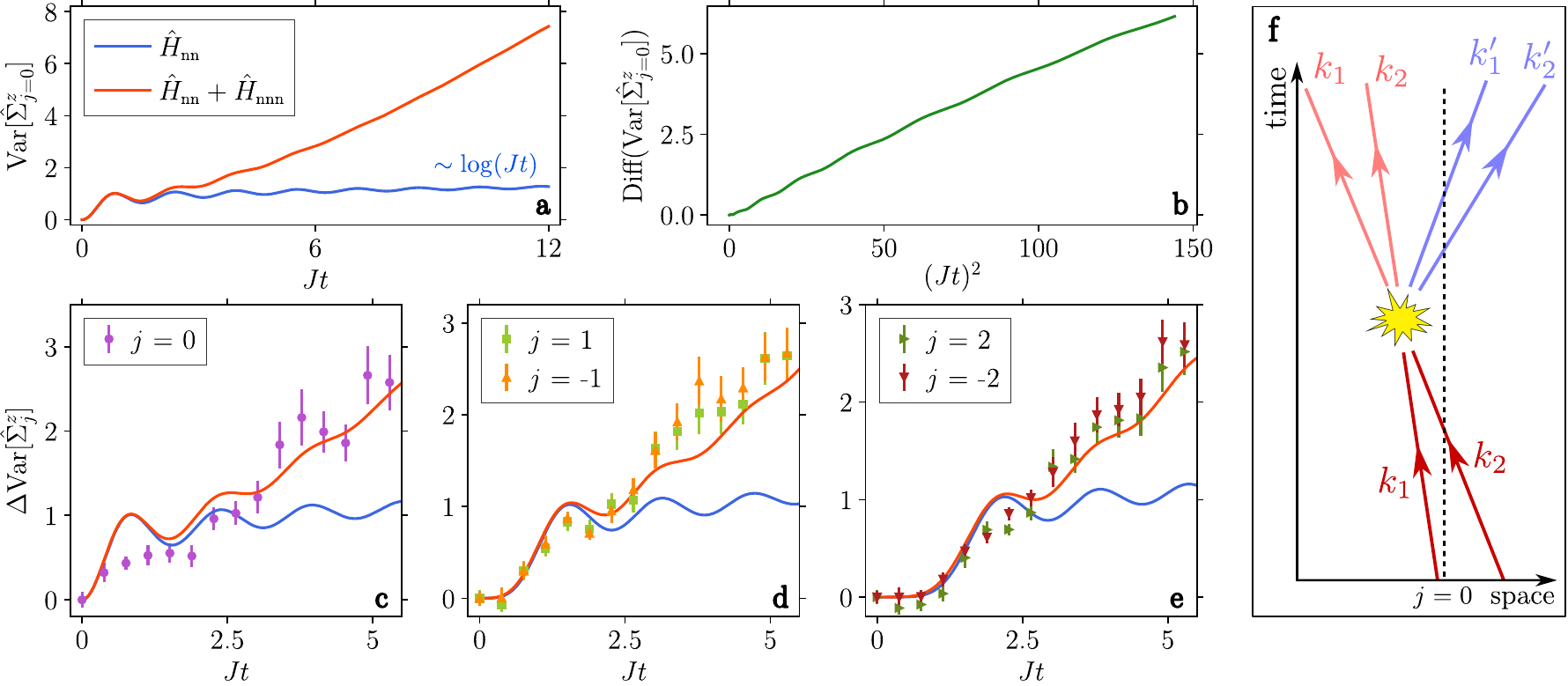}
 \caption{
\textbf{Variance of the half-chain magnetization.} 
\textbf{a}: Numerical simulations of $\text{Var}[\hat \Sigma^z_{j=0}]$ under Hamiltonian evolution with $\hat H_{\rm nn} $ and with $\hat H_{\rm nn} +\hat H_{\rm nnn}$; the chain has length $L=48$.
\textbf{b}: Difference between the two curves shown in \textbf{a}. \textbf{c}, \textbf{d} and \textbf{e}: Experimental data for $\Delta \text{Var}[\hat \Sigma^z_j]$ for several values of $j$ after postprocessing (see the text), and comparison with $\text{Var}[\hat \Sigma^z_{j=0}]$ for the integrable $\hat H_{\rm nn}$ and the non-integrable $\hat H_{\rm nn}+\hat H_{\rm nnn}$.
\textbf{f}: Sketch of the backscattering processes induced by Hamiltonian $\hat H_{\rm nnn}$: a pair of fermionic quasiparticles $(k_1,k_2)$ can be turned into a coherent superposition of pairs of particles propagating in opposite directions, $(k_1,k_2)$ and  $(k'_1,k'_2)$;
eventually they will be on the two different halves of the system, $j\leq 0$ or $j>0$, and let $\hat \Sigma^z_{j=0}$ develop important quantum fluctuations.
 }
 \label{Fig:NoLogGrowth}
\end{figure*}

We now turn to the magnetization dynamics $\langle \hat \sigma^z_j\rangle(t)$ measured on our quantum simulator;  we postselect measurement outcomes by enforcing conservation of the total magnetization: all bit strings whose total magnetization differs from the initial value are discarded, as these events are unambiguously associated with the experimental errors discussed in \emph{Materials and Methods}.
 The spatio-temporal data for $\langle \hat \sigma^z_j\rangle(t)$ shown in
Fig.~\ref{Fig:Magnetization}\textbf{d} exhibit a clear light-cone structure.
Unlike the large-system numerics, a rescaling with $j/(Jt)^{1/z}$ does not reveal a
distinct two-scale (ballistic--diffusive) pattern: the experimental points are equally
consistent both with a ballistic collapse and with a diffusive one, as illustrated in
Figs.~\ref{Fig:Magnetization}\textbf{e} and~\ref{Fig:Magnetization}\textbf{f}. 
In \emph{Materials and Methods} we present a more in-depth study of the transport signatures hidden in the experimental magnetization profile. 
The routine study of the transferred magnetization is compatible with ballistic transport and allows us to fit a dynamical exponent $z = 1.09 \pm 0.03$.
However, we also show that a detailed analysis of the slope of the magnetization profile around the junction yields information reveals an unambiguous signature of the diffusive process that is theoretically expected to take place.
Understanding the difference between these analyses and why one should be more robust than the other is an exciting theoretical perspective.

To assess whether this absence of an unambiguous diffusive broadening can be attributed to experimental imperfections, we perform exact-diagonalization simulations of the same protocol on a finite chain of length $L=14$ evolved with $\hat H_{\rm nn}+\hat H_{\rm nnn}$.
The numerical results (red dots in Fig.~\ref{Fig:Magnetization}\textbf{e}) closely track the experimental data, supporting the conclusion that the dominant limitation is finite size rather than imperfect control. 
We conclude that, at the level of the local magnetization $\langle \hat \sigma^z_j\rangle(t)$, the dynamics accessible in the present experiment are also compatible with a ballistic behavior, consistent with the expectation discussed at the end of the \emph{Introduction}. 

\paragraph{\textbf{Dynamics: Variance of the subsystem magnetization --}}

To make the effect of the breaking of fragile conservation laws more apparent, we consider an integrated (subsystem) magnetization extending from the first site up to site $j$,
and analyze its variance,
\begin{equation}
 \hat \Sigma^z_j =
 \hspace{-0.25cm}
 \sum^j_{m=- \frac{L}{2}+1}
 \hspace{-0.25cm}
 \hat \sigma^z_m
 ; \qquad \text{Var} [\hat \Sigma^z_j ]=
\langle \hat \Sigma^z_j \hat \Sigma^z_j \rangle -
\langle \hat \Sigma^z_j \rangle^2.
\label{Eq:Integrated:Sigma}
\end{equation}
This observable is intrinsically nonlocal: expanding the second moment yields
$\langle (\hat \Sigma^z_j)^2\rangle=\sum_{m=-L/2+1}^{j}\sum_{n=-L/2+1}^{j}
\langle \hat \sigma^z_m \hat \sigma^z_n\rangle$, so that
$\text{Var}[\hat \Sigma^z_j]$ integrates all two-point correlations within the
subsystem.

Figure~\ref{Fig:VarianceIntegrMag}\textbf{a} shows MPS simulations of $\text{Var}[\hat \Sigma^z_j](t)$ for a long chain, $L=48$.
For the value $j=L/2$, we are effectively considering the variance of the total magnetization because we are summing over the entire setup:
it is identically zero because the dynamics conserves the total magnetization $\hat \Sigma^z_{L/2}$, and it is therefore not plotted. As expected, $\text{Var}[\hat \Sigma^z_j](t)$ exhibits a light-cone structure: its growth is ultimately driven by the magnetization dynamics initiated at the junction and then propagated through the chain.

We now turn to the experimental data. 
The postselection described before is here crucial in order to let a clean light-cone pattern appear for $\text{Var}[\hat \Sigma^z_j]$ displayed in
Fig.~\ref{Fig:VarianceIntegrMag}\textbf{b}; a direct evaluation of $\text{Var}[\hat \Sigma^z_j]$ from the measured bit strings produces a very noisy signal.
To obtain a more quantitative characterization of the dynamics, we focus on the
half-chain observable $\text{Var}[\hat \Sigma^z_{j=0}](t)$.
Figure~\ref{Fig:NoLogGrowth}\textbf{a} compares numerical results for the evolution
generated either by the full Hamiltonian $\hat H_{\rm nn}+\hat H_{\rm nnn}$ or by the
nearest-neighbor term $\hat H_{\rm nn}$ alone. When $\hat H_{\rm nnn}$ is included we
observe a rapid, pronounced increase of the variance, in stark contrast to the much
slower $\propto \log t$ growth characteristic of the integrable nearest-neighbor
dynamics~\cite{Eisler2014Surface,Dubail_2017_SciPost, Collura_2020_PRL, Scopa_2021, Ares_2022, Scopa_2022_SciPost, Scopa_2023, Eisler_2025_Scipost}. In the XX case, integrability implies that, inside the light cone, the
state remains locally equivalent to a (locally boosted) Fermi sea, which strongly
constrains the buildup of fluctuations~\cite{Scopa_2023}. In Fig.~\ref{Fig:NoLogGrowth}\textbf{b} we
plot the difference between the two curves; it reveals an initial $\propto t^2$
growth that subsequently crosses over to a slower trend.

We identify the fast growth of $\text{Var}[\hat \Sigma^z_{j=0}](t)$ as a sensitive diagnostic of quasiparticle scattering induced by $\hat H_{\rm nnn}$, which comprises quasiparticle interaction terms.
In particular, the backscattering processes sketched in Fig.~\ref{Fig:NoLogGrowth}\textbf{f}, which are possible because fragile conservation laws are broken, can convert pairs of left-moving quasiparticles into right-moving ones, and create excitations that are the coherent superpositions of quasiparticles moving to the left and to the right.
Ultimately, these quasiparticles will be to the left and to the right of the junction, respectively, making the half-system magnetization subject to quantum fluctuations.
The presence (or absence) of such bidirectionally propagating excitations is precisely what the integrated variance
$\text{Var}[\hat \Sigma^z_{j=0}]$ is designed to capture.

We next turn to the experimental data. Unlike in the numerics, we observe a finite
variance already at the initial time, $\text{Var}[\hat \Sigma^z_j](t=0)\neq 0$.
We attribute this offset to imperfect preparation of the initial state and to
readout errors (see \emph{Materials and Methods} for details). Since these
imperfections contribute at all times and are not the focus of our analysis---which targets fluctuations generated by the coherent unitary dynamics---we subtract the
initial value and consider the variance growth
$\Delta\text{Var}[\hat \Sigma^z_j](t)
=\text{Var} [ \hat \Sigma^z_{j} ](t)-\text{Var} [ \hat \Sigma^z_{j} ](t=0)$. 
Figures~\ref{Fig:NoLogGrowth}\textbf{c}--\ref{Fig:NoLogGrowth}\textbf{e} show
$\Delta\text{Var}[\hat \Sigma^z_j](t)$ for several choices of $j$.
We find a clear increase with time that is incompatible with the logarithmic growth
expected for the integrable nearest-neighbor dynamics.
This shows that magnetization fluctuations, which have been already
explored in other quantum-simulation platforms and in different contexts~\cite{Jacqmin2011Sub,Wienand2024Emergence, Rosenberg_2024}, and in some cases are even related to entanglement entropies~\cite{Klich_2006, Song_2010, Song_2011, Rachel2012Detecting, Calabrese_2012}, provide a
particularly sensitive probe of weak breaking of fragile conservation laws that characterize the dynamics under $\hat H_{\rm nn}$.

\paragraph{\textbf{Dynamics: String operator --}}

Besides variances, which quantify fluctuations, other nonlocal operators play a central role in many-body physics. A notable example, particularly in the context of topological order, is provided by string operators, which can reveal hidden order in bosonic gases or spin chains~\cite{DallaTorre_2006, Rossini_2012, LESELUC_2019, Moegerle_2025} and has also been measured in cold-atom experiments using a quantum-gas microscope~\cite{Endres_2011, Hilker_2017}. While intrinsically nonlocal in spin language, they can represent local observables when framed in terms of fermionic quasiparticles, but only when the full physical setting (state and Hamiltonian) is invariant under the corresponding symmetry~\cite{PerezGarcia_2008, Fagotti2024Nonequilibrium}. Here we show that such operators provide an equally sharp diagnostic of breaking of fragile conservation laws, displaying a pronounced qualitative difference with respect to the integrable case already at short times. 

To this end, we introduce the string operator
\begin{equation}
\hat P^z_j =
\hspace{-0.25cm}
\bigotimes^j_{m=-\frac L2+1}
\hspace{-0.25cm}
\hat \sigma^z_m = (-1)^{ \sum_m^j \frac12\left(\hat \sigma^z_m+1 \right)}.
\end{equation}
The second equality makes explicit that $\hat P_j^z$ measures, up to an inessential prefactor, the parity of the magnetization accumulated on the interval $[-L/2+1,j]$. Equivalently, it can be viewed as a particular instance of full-counting statistics (namely, the generating function evaluated at the special point that selects the parity sector).

\begin{figure*}[t]
\centering
\includegraphics[width=0.9\textwidth]{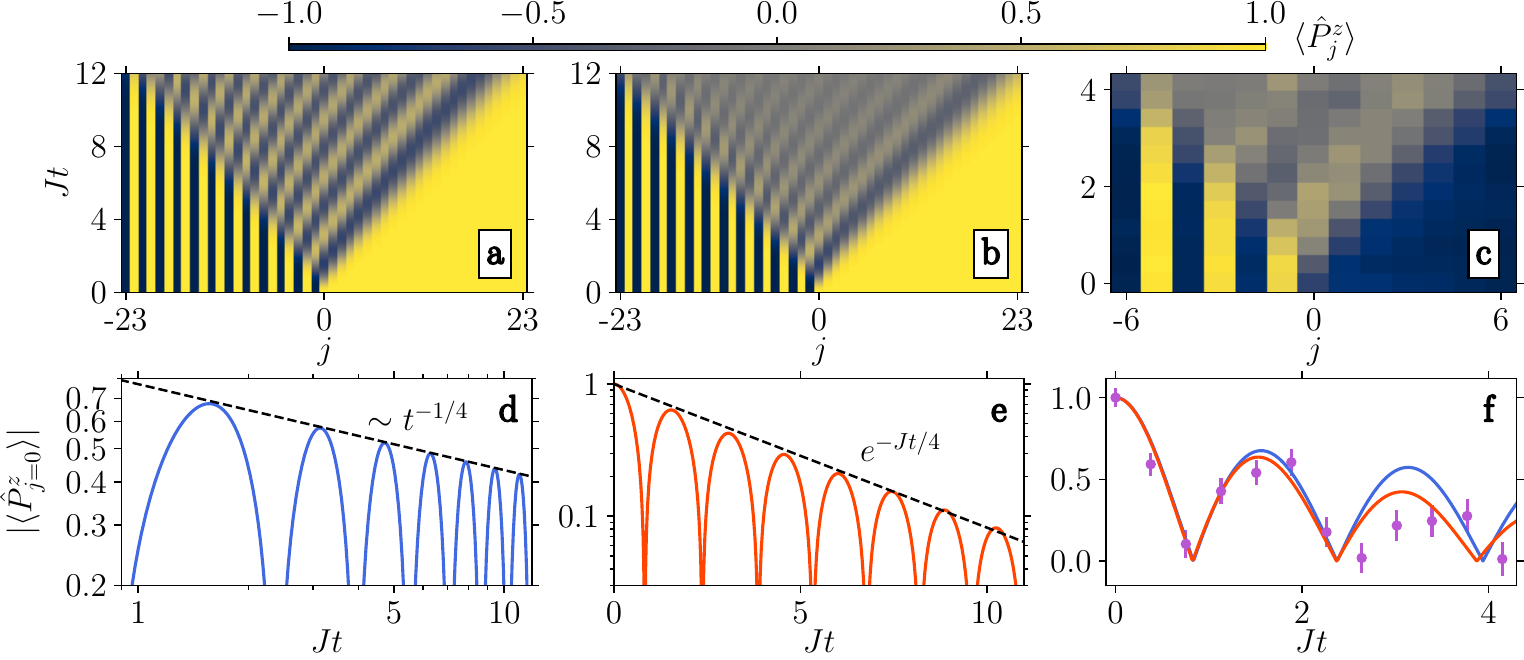}
 \caption{
\textbf{String operator.}  \textbf{a} and \textbf{b}: Numerical space-time profiles of $\langle \hat P^z_j \rangle(t)$ for evolution under $\hat H_{\rm nn}$ and $\hat H_{\rm nn}+\hat H_{\rm nnn}$, respectively.
\textbf{c}: Experimental space-time profile of $\langle \hat P^z_j \rangle(t)$.
 \textbf{d} and \textbf{e}: Numerical time traces of $\langle \hat P^z_{j = 0} \rangle(t)$ for evolution under $\hat H_{\rm nn}$ and $\hat H_{\rm nn}+\hat H_{\rm nnn}$, respectively.
\textbf{f}: Experimental data for $\langle \hat P^z_{j = 0} \rangle(t)$, rescaled to start from $1$.
}
 \label{Fig:String:Operator}
\end{figure*}

We begin by presenting in Figs.~\ref{Fig:String:Operator}\textbf{a} and \ref{Fig:String:Operator}\textbf{b} numerical simulations of $\langle \hat P^z_j\rangle(t)$ for time evolution generated by $\hat H_{\rm nn}$ and by $\hat H_{\rm nn}+\hat H_{\rm nnn}$, respectively. In both cases the dynamics display a clear light-cone structure. The strongly staggered pattern at $t=0$, alternating between $+1$ and $-1$, reflects the presence of sites which are in the $\ket{\downarrow}$ state; it is natural to interpret these as \emph{hole} quasiparticles with respect to the fully polarized reference state $\ket{\uparrow}$. Each hole carries one quantum of negative magnetization, and in the initial domain-wall state $\ket{\Psi_{\rm DW}}$ holes occupy the left half of the chain with unit filling. As time evolves, the staggered structure progressively spreads into the right half, which initially contains no holes. Tracking the resulting trajectories therefore provides a direct real-time picture of hole propagation across the junction. While the two space-time plots are qualitatively similar, the pattern is visibly more persistent in the integrable case.

To quantify this difference, we focus on the half-chain string operator and study
the time evolution of $\langle \hat P^z_{j=0}\rangle(t)$. The numerical results shown
in Figs.~\ref{Fig:String:Operator}\textbf{d} and~\ref{Fig:String:Operator}\textbf{e} reveal a slow algebraic
decay $\propto t^{-1/4}$ for the integrable dynamics generated by $\hat H_{\rm nn}$,
whereas adding $\hat H_{\rm nnn}$ leads to a markedly faster decay compatible with
$\propto e^{-Jt/4}$. The algebraic behaviour indicates that, in the integrable case,
the hole quasiparticles retain long-lived quantum coherence, while in the
nonintegrable case this coherence is rapidly degraded by interactions and
backscattering. 
Thus, the string operator provides a sharp diagnostic, clearly
separating the integrable regime of effectively free fermions from the interacting
regime in which quasiparticle scattering is no longer purely elastic.

We now turn to the experimental observations. Figure~\ref{Fig:String:Operator}\textbf{c} shows the measured spatio-temporal profile of $\langle \hat P^z_j\rangle(t)$, which exhibits a clear blurring of the striped pattern. We interpret this loss of contrast as a signature of quasiparticle scattering induced by the long-range couplings. To make this statement quantitative, Fig.~\ref{Fig:String:Operator}\textbf{f} reports the half-chain signal $\langle \hat P^z_{j=0}\rangle(t)$ extracted from the experiment, which is compatible with an exponential decay. 
Nevertheless, the figure also highlights that for such short time scales, it is very difficult to distinguish an exponential decay from an algebraic one.
Taken together, the results in Fig.~\ref{Fig:String:Operator} show that string operators provide an experimentally accessible probe of the weak breaking of the fragile conservation laws of $\hat H_{\rm nn}$ induced by dipolar interactions in the Rydberg-atom platform.

\paragraph{\textbf{A theoretical model} ---}

\begin{figure}[b]
\includegraphics[width=0.65\columnwidth]{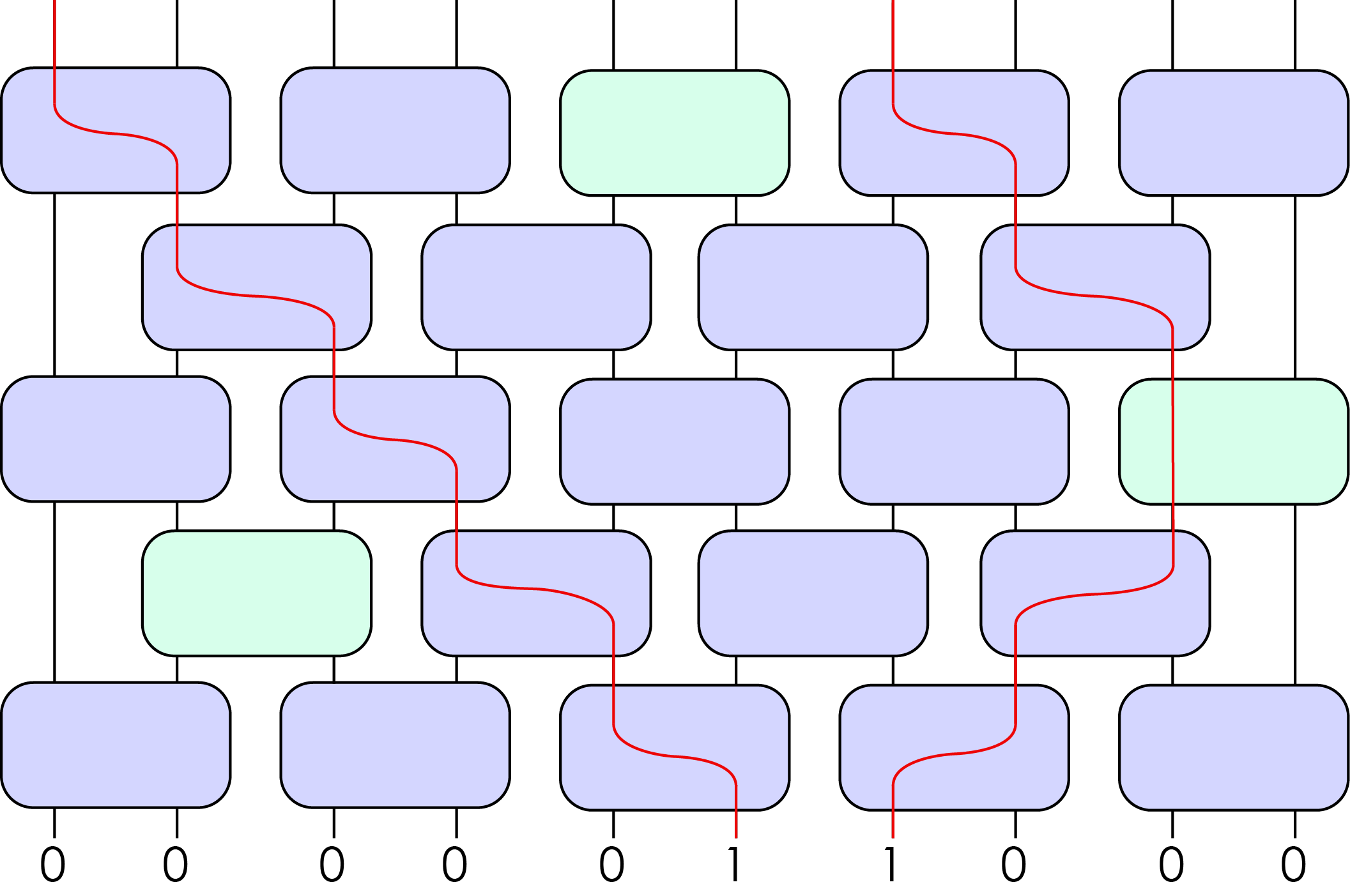}
 \caption{
 \textbf{The classical statistical cellular automaton.} \textit{Swap} gates are shown in violet and \textit{identity} gates in green. The red line highlights the right- and left-moving excitations supported by the \textit{swap} gates, while the \textit{identity} gates backscatter them.
 }
 \label{Fig:Cellular:Sketch}
\end{figure}

From a theoretical perspective, it is important to relate the rapid growth of
$\text{Var}[\hat \Sigma^z_{j=0}](t)$ to the backscattering processes generated by
the integrability-breaking term $\hat H_{\rm nnn}$ and sketched in Fig.~\ref{Fig:NoLogGrowth}\textbf{f}. Remarkably, the key qualitative
features of this mechanism can already be reproduced by a simple classical model,
which provides an intuitive and numerically tractable description of the effect. We
therefore use this classical setting to isolate the relevant physics, in the same spirit of Ref.~\cite{bd-17}.

\begin{figure*}[t]
\centering
\includegraphics[trim=4cm 0 4cm 0,clip,width=\textwidth]{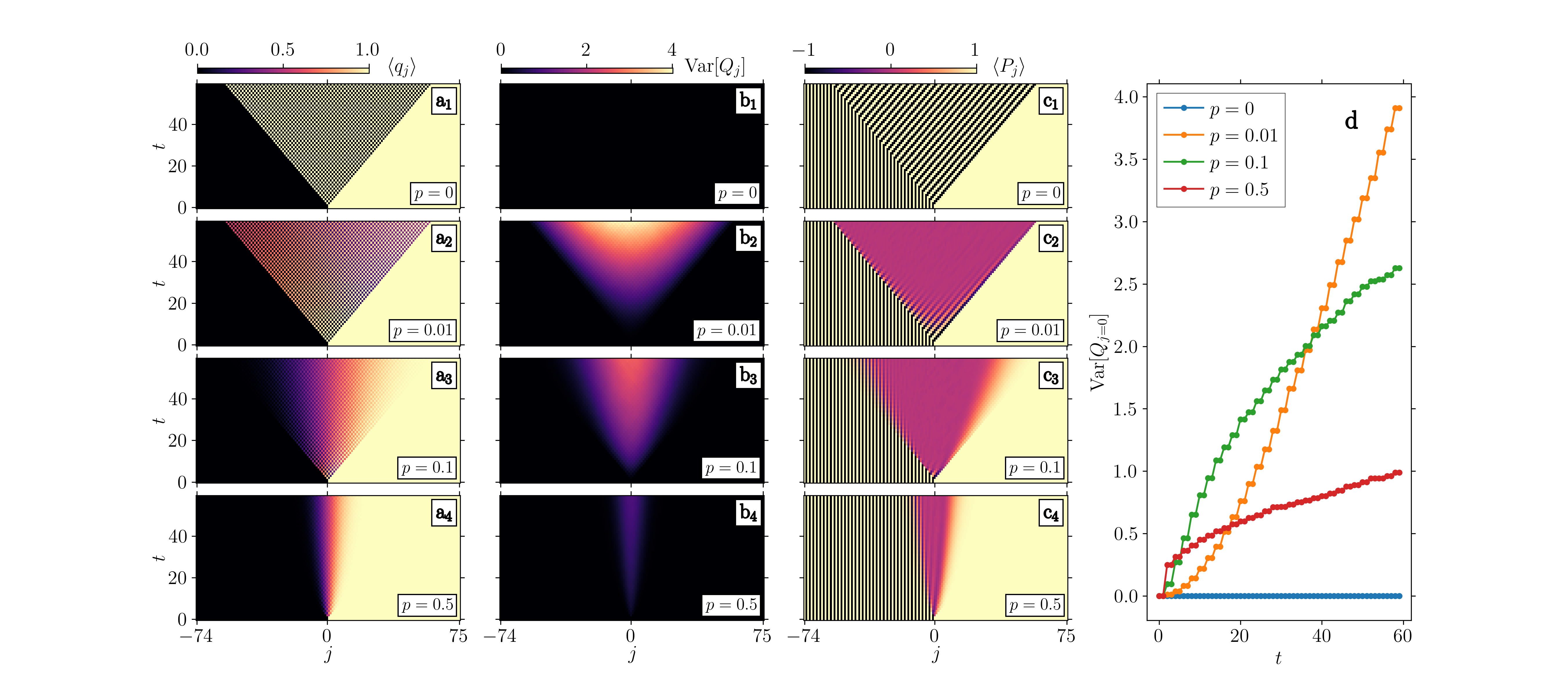}
 \caption{
\textbf{Weak integrability breaking in classical stochastic dynamics.}
\textbf{a}: Average particle density $\langle q_j\rangle(t)$ for different values of $p$, starting from a domain-wall configuration. We consider a system of size $L=150$ and average over $N=10^4$ realizations. The light cone, sharp at $p=0$, progressively melts for $p\neq 0$, crossing over toward diffusive dynamics; for large values (e.g., $p=0.5$) only diffusive behavior is visible.
\textbf{b}: Space-time evolution of the variance of the number of particles in the interval $[-L/2+1,j]$, $\text{Var}[ Q_j](t)$. For $p=0$ the particle number does not fluctuate, while for small $p$ strong fluctuations develop and become less pronounced as $p$ is increased.
\textbf{c}: Space-time evolution of the average string observable, $\langle P_j \rangle (t)$. The plots clearly show the melting of the diagonal features associated with hole quasiparticles propagating from the left half into the right half.
\textbf{d}: Half-chain variance $\text{Var}[Q_{j=0}](t)$ as a function of time $t$ for different values of $p$. At late times, the largest variance is reached for the smallest nonzero value of $p$.
 }
 \label{Fig:Authomaton}
\end{figure*}

We consider a one-dimensional stochastic cellular automaton whose configurations are
bit strings with entries in $\{0,1\}$. The dynamics is defined by local two-site
update rules, represented as classical gates acting on nearest-neighbor pairs. The
gates are applied sequentially in the standard brickwork pattern, with alternating
even and odd layers~\cite{Fisher-23}; a schematic is shown in
Fig.~\ref{Fig:Cellular:Sketch}. On each updated bond the gate is chosen randomly as
follows: (i) with probability $1-p$ the two bits are exchanged (a \emph{swap} gate),
and (ii) with probability $p$ the bits are left unchanged (an \emph{identity} gate).

At $p=0$ the evolution is purely ballistic. Interpreting the ``$1$''s as particles,
one immediately finds stable right- and left-moving quasiparticles propagating with
velocity $\pm 1$. For small but nonzero $p\ll 1$, the dominant effect is the
appearance of backscattering events that convert right-movers into left-movers, and
vice versa. Activating this scattering channel breaks the strictly ballistic
dynamics of the $p=0$ limit; at sufficiently long times, repeated backscattering is
therefore expected to produce diffusive transport. This makes the
automaton a minimal setting in which to analyze the crossover from ballistic to
diffusive behaviour~\cite{Medenjak-17,Klobas-18,Krajnik-25}, and we will use it as a
qualitative proxy for the quantum dynamics discussed above.
While a direct correspondence between this classical model and the quantum dynamics discussed above should be treated with caution, the model can nevertheless serve as a qualitative guide: in the spirit of the Fermi golden rule, one may generically expect $p$ to be of the same order as $( 1/8 )^2$, corresponding to the squared strength of the perturbation $\hat H_{\rm nnn}$ relative to the integrable Hamiltonian $\hat H_{\rm nn}$.

We consider the particle density at site $j$ and time $t$, denoted by $q_j(t)$, and
its average over stochastic realizations, $\langle q_j\rangle(t)$. We study the
evolution from a domain-wall initial condition $(\ldots 000111 \ldots)$. The
resulting profiles are shown in Fig.~\ref{Fig:Authomaton}\textbf{a} for several
values of $p$. For $p=0$, a sharp light cone develops, reflecting ballistic scaling
$j\sim t$. Inside the cone, the system forms a \emph{nonequilibrium steady state}
which, in the large-system and long-time limit, becomes a scaling function of
$j/t$ (see \emph{Materials and Methods}), mirroring the integrable quantum dynamics
generated by $\hat H_{\rm nn}$. For $p\neq 0$ the light cone is progressively
degraded and, at sufficiently long times, a parabolic region with width growing as
$j\sim\sqrt{t}$ becomes apparent. This is a clear signature of diffusion, analogous
to the diffusive scaling observed in the rescaled quantum data for
$\langle \hat \sigma^z_j\rangle(t)$ in Fig.~\ref{Fig:Magnetization}\textbf{c}.

In direct analogy with Eq.~\eqref{Eq:Integrated:Sigma}, we also consider the
integrated particle number in the interval $[1,j]$, defined as
$Q_j=\sum_{m\le j} q_m$, and analyze its variance $\text{Var}[Q_j](t)$.
Figure~\ref{Fig:Authomaton}\textbf{b} shows $\text{Var}[Q_j](t)$ as a function of
space and time. In the deterministic limit $p=0$ the variance vanishes identically,
in particular $\text{Var}[Q_{j=0}](t)=0$, reflecting the absence of stochastic
fluctuations. It is therefore tempting to view the slow $\sim\log t$ increase of the
half-chain variance $\text{Var}[\hat \Sigma^z_{j=0}](t)$ under $\hat H_{\rm nn}$ as
a genuinely quantum analogue of this deterministic behavior.

For $p\neq 0$, the support of $\text{Var}[Q_j](t)$ follows the same combined
ballistic--diffusive structure seen in the density profile $\langle q_j\rangle(t)$.
More strikingly, the half-chain variance $\text{Var}[Q_{j=0}](t)$, shown in
Fig.~\ref{Fig:Authomaton}\textbf{d}, exhibits a counterintuitive trend: as expected,
a smaller scattering probability $p$ yields a slower initial increase, but at
intermediate times the growth becomes more pronounced the smaller $p$ is.
Thus, weak integrability breaking leads to a rapid buildup of half-chain fluctuations, whereas
in the integrable limit the variance remains strongly suppressed, a phenomenology
closely reminiscent of our results in the chain of Rydberg atoms. Indeed, at short times, both in the classical and in the quantum setting, half-chain variances display an initial $\sim t^2$ growth.

Finally, to tighten the analogy with the quantum data, we also consider a string
observable, $P_j = e^{i\pi Q_j}$, which measures the parity of the number of particles in the interval
$[-L/2+1,\,j]$. The corresponding data, shown in Fig.~\ref{Fig:Authomaton}\textbf{c},
exhibit, in the deterministic limit $p=0$, a pronounced diagonal striped pattern
associated with the ballistic motion of hole quasiparticles from left to right. As
$p$ is increased, the overall light-cone structure remains visible, but the striped
pattern loses contrast and is washed out exponentially in time (data not shown). For
larger values of $p$, the light cone itself eventually fades, giving way to a purely
diffusive, parabolic broadening.

To clarify the phenomenology observed in our numerics, it is useful to analyze the
automaton in the dilute-scattering regime $p\ll 1$ and adopt a coarse-grained
hydrodynamic viewpoint in continuous space and time. As discussed in more detail in
\emph{Materials and Methods}, this limit can be mapped onto a \emph{run-and-tumble}
process, which is a standard model in classical statistical physics~\cite{Angelani-15,bmrs-20,Jode-23}.
Within this description one obtains a closed evolution equation for the coarse-grained
particle density $q(x,t)$, the \emph{telegrapher equation},
\begin{equation}
-\partial_t^2 q + \partial_x^2 q = 2\Gamma\,\partial_t q,
\label{Eq:Telegrapher}
\end{equation}
where $\Gamma$ is the tumbling (velocity-randomization) rate and plays the same role
as the backscattering probability $p$ in the microscopic automaton.

Equation~\eqref{Eq:Telegrapher} is known to exhibit two distinct regimes.
At short times, the damping term is negligible compared with the inertial term,
$2\Gamma\,\partial_t q \ll \partial_t^2 q$, and the dynamics is effectively ballistic,
with a light cone propagating at unit velocity. At late times, the inertial term
becomes subleading, $\partial_t^2 q \ll 2\Gamma\,\partial_t q$, and the evolution
reduces to the diffusion equation with diffusion constant $D=\Gamma^{-1}$. The
increase of $D$ as $\Gamma$ decreases admits a simple interpretation: rare tumbling
events allow particles to travel farther between velocity reversals, thereby
enhancing spatial spreading. This ballistic-to-diffusive crossover is precisely what
we observe in the cellular-automaton density profiles in Fig.~\ref{Fig:Authomaton}\textbf{a}.

Analytical results are also available for the fluctuations of the integrated number
$Q_j$ in the run-and-tumble model~\cite{Jode-23}. In particular, the half-chain
variance obeys the scaling $\text{Var}[Q_j](t)\sim \Gamma t^2$ for
$t\ll \Gamma^{-1}$ and $\text{Var}[Q_j](t)\sim \sqrt{t/\Gamma}$ for
$t\gg \Gamma^{-1}$. This is noteworthy for two reasons. First, it matches the trends
observed for the cellular automaton in Fig.~\ref{Fig:Authomaton}\textbf{d}. Second,
it reproduces the characteristic short-time growth $\sim t^2$ that accompanies weak
breaking of conservation laws and that we identified in the quantum simulations
(Fig.~\ref{Fig:VarianceIntegrMag}\textbf{b}). Finally, the late-time behaviour is
non-analytic as $\Gamma\to 0^+$: although $\text{Var}[Q_j](t)$ vanishes identically
at $\Gamma=0$ (the strictly ballistic, deterministic limit), for any $\Gamma>0$ it
eventually grows as $\sqrt{t/\Gamma}$. This singular dependence on the
integrability-breaking rate appears to be a robust fingerprint of weak violations of
conservation laws, from classical stochastic dynamics to the quantum setting.

{\bf Discussion}

Our work reports the first experimental measurements of the effects of weak breaking of fragile conservation laws in a small quantum simulator. The physical consequences of integrability in quantum spin chains, most notably ballistic transport, have already been probed in several quantum-simulation platforms~\cite{Kinoshita2006, Langen2015Science, Schemmer_2019, Malvania2021, Wei2022, Dubois_2024, Schuettelkopf2025}. 
Likewise, strongly nonintegrable setups such as spin ladders have provided clear signatures of diffusive transport~\cite{Wei2022, Wienand2024Emergence}. 
The present study sits in an intermediate regime, in which the robust conservation laws of the underlying integrable model remain effectively conserved on the time scales we probe, which has also been explored with several quantum simulators~\cite{Gring2012, Langen_2016, Moller_2024};
we have shown that the breakdown of fragile conservation laws leaves clear signatures at short times in small setups.

Experimentally, the magnetization profile is compatible with both ballistic and diffusive rescaling: while there are indications that transport may become diffusive in larger systems at later times, similarly compelling signatures of ballistic behavior are also present. 
In this respect, our numerical simulations on significantly larger systems provide evidence for a ballistic light cone, within which diffusive transport develops in the vicinity of the junction.
In contrast, we have provided numerical and experimental evidence that the variance of the subsystem magnetization grows in time in a fashion incompatible with the propagation of noninteracting fermionic quasiparticles. Our results show that interactions between the fermionic quasiparticles carrying magnetization generate delocalized excitations that are coherent superpositions of left- and right-movers. Following this line of thought, we have also demonstrated that other observables represented by nonlocal string operators are sensitive probes of the breaking of fragile conservation laws induced by weak integrability breaking in Rydberg chains.

To further elucidate the underlying mechanism, we analyzed a classical stochastic cellular automaton with an integrable point and a well-controlled regime of weak integrability breaking. Although this model is classical, and the comparison to quantum dynamics must be made with care, it reproduces several qualitative features of our experimental and numerical observations. Its main advantage is conceptual: it provides an intuitive, numerically accessible setting in which the impact of the breakdown of fragile conservation laws under weak perturbations can be explored. This successful comparison suggests that a fully quantum theoretical description of the behavior of the two nonlocal quantities we measure, $\text{Var} [\hat \Sigma^z_j]$ and $\langle \hat P^z_j \rangle(t)$, should be within reach, even if far from elementary. In particular, identifying how these observables scale with the small parameter controlling weak integrability breaking remains a key open problem. It is not obvious that existing approaches, essentially based on Fermi–Golden–Rule-type treatments of weak perturbations, can be straightforwardly applied to this setting.

More broadly, our results highlight the experimental relevance of the physics of weak breaking of conservation laws in quantum many-body systems. In classical mechanics, the KAM theorem provides the conceptual foundation for treating the regime of weak integrability breaking as a field in its own right, distinct from both perfectly integrable and strongly chaotic dynamics. An analogous, fully developed framework is still missing in the many-body context, let alone the quantum case. While the literature contains many perturbative constructions that attempt to deform conserved charges, these are typically truncated at low order~\cite{Jung_2006, Bargheer_2008, Bargheer_2009, Brandino_2015, Szasz_2021, Kurlov2022, Surace_2023, Orlov_2023, Vanovac_2024}. Extending such techniques to situations in which the perturbative series can be meaningfully resummed, closer in spirit to the KAM theorem, would allow one to make definite statements about late-time dynamics under weak integrability breaking. Our work offers concrete experimental and numerical benchmarks for such a future quantum theory of weak integrability breaking, and underscores the central role of fluctuations and nonlocal observables in diagnosing the fragility of conservation laws.

\textbf{Materials and Methods}

\paragraph{\textbf{Experimental design ---}}
The implementation of the dipolar XX Hamiltonian is described in our previous works~\cite{Chen2023, Bornet2023,2025_Chen,Emperauger_2025}. 
The pseudo-spin states $\ket{\uparrow} = \ket{60S_{1/2},m_J=1/2}$ and  $\ket{\downarrow} = \ket{60P_{1/2},m_J=-1/2}$ can be coupled 
by microwave at 16.7~GHz. We apply a $\sim45$~G quantization magnetic field perpendicular to the array, to keep isotropic interactions and isolate the $\ket{\uparrow}-\ket{\downarrow}$ transition from irrelevant Zeeman sublevels.

First, all the atoms are excited to the Rydberg state $\ket{\uparrow}$ by a stimulated Raman adiabatic passage (STIRAP) 
with 421~nm and 1013~nm lasers. Before initializing the system in $\ket{\Psi_{\text{DW}}}$,
we apply a global resonant microwave $\pi$-pulse to transfer all the atoms in  $\ket{\downarrow}$ state~\cite{Chen2023}, with a Rabi frequency $\Omega = 2\pi \times 19.2$~MHz. Then, we apply an addressing light-shift ($\sim 39$~MHz) on half of the atoms and, simultaneously, a weaker microwave $\pi$-pulse ($\Omega = 2\pi \times 15.1$~MHz) to transfer only the non-addressed atoms back to $\ket{\uparrow}$ while keeping  the addressed atoms in $\ket{\downarrow}$, leading to the state $\ket{\Psi_{\text{DW}}}$.

Several sources of state preparation and measurement (SPAM) error contribute to affecting the observed magnetization and variance. We estimate that the Rydberg excitation process has a efficiency of $98\%$; thus, a fraction $\eta = 2\%$ of the atoms remain in the state $|g\rangle$ after Rydberg excitation and are excluded from the following dynamics. 
These uninitialized atoms are read as a spin $\left\vert \uparrow \right\rangle$ at the end of the sequence.
Applying the microwave $\pi$-pulse simultaneously with the addressing laser, we find the non-addressed atoms in the right-half of the chain have an efficiency $ \eta_{\text{non-add}}$ to be transferred back to $\ket{\uparrow}$, and the addressed atoms in the left half have $\eta_{\text{add}}$ to be transferred  to $\ket{\uparrow}$. 
With $\sim 39~$MHz addressing light-shift, we find
$ \eta_{\text{non-add}} = 95\%$ and $\eta_{\text{add}} = 1\%$.

Due to the finite efficiency of  the readout sequence, an atom in $\ket{\uparrow}$ (resp.~$\ket{\downarrow}$) has a non-zero probability $\epsilon_{\uparrow}$ (resp.~$\epsilon_{\downarrow}$) to be detected in the wrong state~\cite{Chen2023}. 
The main contributions to $\epsilon_{\uparrow}$ are the finite efficiency $1-\eta_{\text{dx}}$ of the de-excitation pulse and the probability of loss $\epsilon$ due to collisions with the background gas. 
As for $\epsilon_{\downarrow}$, the main contribution is the $\ket{\downarrow}$ Rydberg state radiative lifetime. A set of calibrations leads, to first order, to $\epsilon_{\uparrow} \simeq \eta_{\text{dx}} + \epsilon = 1.5\% + 1.0\% = 2.5\%$ 
and $\epsilon_{\downarrow} = 1.0\%$.
Considering both preparation and readout errors, we find the experimental prepared initial state has averaged recapture probabilities of $91.9\%$ for non-addressed atoms and $4.9\%$ for addressed atoms, corresponding to the measured magnetizations at $t = 0$ in Fig.~\ref{Fig:Magnetization}\textbf{d}.

\paragraph{\textbf{Integrability in quantum spin chains and the mapping to fermionic quasiparticles ---}}

Integrability in quantum many-body systems is a subtle notion, and no single
definition is universally accepted~\cite{Caux_2011}. In this work we adopt a
perspective based on \emph{(pseudo)local conserved charges}~\cite{Ilievski2016Quasilocal,Doyon2017Thermalization},
which in the present setting reduces to the more concrete notion of
\emph{extensive integrals of motion with local densities}~\cite{Essler_2016}. Such conserved quantities,
often referred to as \emph{conservation laws}, typically take the form
$\hat Q=\sum_j \hat q_j$, where the density $\hat q_j$ has finite spatial support
around site $j$ (not necessarily a single-site operator).

This locality requirement has an immediate conceptual advantage: it excludes from
the discussion the projectors onto individual Hamiltonian eigenstates, which are
indeed conserved but are highly nonlocal and, in the absence of finite-size effects,
do not constrain the dynamics of local observables. Within this viewpoint, an
integrable quantum system is characterized by the presence of an \emph{extensive} set of
independent local conservation laws, a number that grows proportionally with system
size (for a one-dimensional chain, with the number of sites $L$). Exploiting only
this extensive family of charges has proven sufficient to account for equilibration
in a broad class of integrable models via the generalized Gibbs ensemble~\cite{Rigol_2007, Vidmar_2016, Essler_2016}, including
paradigmatic free-fermion systems such as the XX chain and the transverse-field Ising
model (see for instance some early works in Refs.~\cite{Barthel_2008, Rossini_2009, Fagotti_2011}).

The mapping of $\hat H_{\rm nn}$ and $\hat H_{\rm nnn}$ to spinless fermions is
performed via the Jordan--Wigner transformation~\cite{JordanWigner1928, Sachdev2011}. We introduce
real-space fermionic operators $\hat c_j^{(\dagger)}$ and their Fourier modes
$\hat c^{(\dagger)}(k)$ (with $k$ in the first Brillouin zone), both obeying the
canonical anticommutation relations. This yields
\begin{subequations}
\begin{align}
 \hat H_{\rm nn} =& -\hbar J\sum_j \hat c_{j+1}^\dagger \hat c_j  + \text{H.c.} \nonumber\\
 &=-2\hbar J \int_{-\pi}^\pi\frac{dk}{2\pi} \cos k
 \hat c^\dagger(k) \hat c(k); \\
 \hat H_{\rm nnn} =& - \frac 18 \hbar J \sum_j \hat c_{j+2}^\dagger \hat c_j  -2 \hat c_{j+2}^\dagger \hat n_{j+1}
 \hat c_j   + \text{H.c.}
 \, .\label{Eq:Fermi:nnn}
\end{align}
\end{subequations}
The nearest-neighbor Hamiltonian $\hat H_{\rm nn}$ is quadratic and therefore
describes noninteracting fermions. The $r=2$ term $\hat H_{\rm nnn}$ contains a
quadratic hopping contribution, which renormalizes the single-particle dispersion,
and a genuinely interacting (density-dependent hopping) contribution, which breaks
the free-fermion integrable structure.

Within the viewpoint adopted in this article, the integrability of $\hat H_{\rm nn}$
is made explicit by constructing an extensive family of commuting charges with local
densities. Since $\hat H_{\rm nn}$ is diagonal in momentum space, all mode
occupations $\hat n(k)=\hat c^\dagger(k) \hat c(k)$ are conserved. Taking Fourier
moments of $\hat n(k)$ yields
\begin{equation}
\begin{aligned}
 \hat Q_m^{(1)} = &\int_{-\pi}^\pi\frac{dk}{2\pi} \cos(mk) \hat c^\dagger(k) \hat c(k)\\
 \hat Q_m^{(2)} =& \int_{-\pi}^\pi\frac{dk}{2\pi} \sin(mk) \hat c^\dagger(k) \hat c(k);
\end{aligned}
\end{equation}
which commute pairwise. In real space these operators read~\cite{Grabowski1995Structure}
\begin{equation}
\begin{aligned}
 \hat Q_m^{(1)} =& \, \frac{1}{2}\sum_j \left( \hat c^\dagger_j \hat c_{j+m} + \hat c^\dagger_{j+m} \hat c_{j} \right);
 \\
 \hat Q_m^{(2)} =& \, \frac{i}{2} \sum_j \left( \hat c^\dagger_j \hat c_{j+m} - \hat c^\dagger_{j+m} \hat c_{j} \right),
\end{aligned}
\end{equation}
and are therefore extensive with densities supported on a finite range $m$. Up to
inessential prefactors, $\hat Q^{(1)}_{m=0}$ is the total magnetization,
$\hat Q^{(1)}_{m=1}$ coincides with $\hat H_{\rm nn}$ (energy), and $\hat Q^{(2)}_{m=1}$
is the magnetization current. We caution that the XX chain (more generally, the free-fermion XY family) admits
additional families of local conserved charges beyond the translation-invariant ones,
originating from degeneracies of the single-particle spectrum~\cite{Fagotti2014On}. These extra (typically
staggered/non-translation-invariant) conservation laws become relevant when
one-site translation symmetry is explicitly broken, for instance by dimerization.
Our initial state does not couple to these
additional sectors in a significant way.

If one retains only the quadratic part of $\hat H_{\rm nnn}$ in
Eq.~\eqref{Eq:Fermi:nnn}, the model remains free and $\hat n(k)$ is still conserved;
consequently, the charges $\hat Q_m^{(i)}$ remain integrals of motion (although their
physical interpretation changes, since the energy current and higher currents are
modified by the altered dispersion). This structure is instead broken once the
quartic term in Eq.~\eqref{Eq:Fermi:nnn} is included, as it induces scattering
between fermionic quasiparticles and destroys the conservation of $\hat n(k)$ and of
the associated local charges.

\paragraph{\textbf{Ballistic domain-wall dynamics under $\hat H_{\rm nn}$} ---}

Focusing, for simplicity, on $\hat H_{\rm nn}$, the fermionic quasiparticles created by $\hat c^\dagger(k)$ carry one quantum of magnetization, have dispersion $\varepsilon(k)\propto \cos k$, and propagate with group velocity $v(k)=\partial_k\varepsilon(k)\propto \sin k$. The domain-wall state $\ket{\Psi_{\rm DW}}$ prepares two macroscopically distinct regions: the left half is empty of fermions, while the right half is completely filled.
A convenient way to formulate the ensuing dynamics, and to make contact with hydrodynamic descriptions, is through a phase-space representation in terms of the coarse-grained mode occupation $n(x,k,t)\equiv \langle \hat n(k)\rangle_{x,t}$ for $k \in (-\pi, \pi]$. At $t=0$, the domain wall corresponds to the step profile $n(x,k,0)=\theta(x)$. For the free-fermion dynamics generated by $\hat H_{\rm nn}$, $n(x,k,t)$ obeys the ballistic advection equation $\partial_t n + v(k)\partial_x n=0$, hence $n(x,k,t)=n(x/t,k)=\theta\bigl(x/t-v(k)\bigr)$, which provides an exact ballistic scaling form. This result is the key input behind
the ballistic prediction for the magnetization profile reported in the
\emph{Results}, obtained by integrating $n(x/t,k)$ over $k$.

The next-to-nearest-neighbor term $\hat H_{\rm nnn}$ spoils this picture by inducing
interactions between fermionic quasiparticles. A number of theoretical frameworks have
been developed to treat such weak integrability breaking perturbatively, typically by
deriving generalized kinetic (Boltzmann-type) equations in which the collision
integral is evaluated at leading order using time-dependent perturbation theory and
Fermi’s golden rule~\cite{Bastianello_2021,Mallayya2021Prethermalization}. At present,
however, it is not settled which theoretical approach could provide an accurate
quantitative prediction for the domain-wall dynamics under $\hat H_{\rm nn}+\hat
H_{\rm nnn}$ in the experimentally relevant time window.

\paragraph{\textbf{Fragile conservation laws of $\hat H_{\rm nn}$} ---}

Under weak integrability breaking, a subset of the local conservation laws of $\hat H_{\rm nn}$ can be continuously deformed into quasilocal operators that remain approximately conserved up to times scaling as the square of the inverse perturbation strength (see also Ref.~\cite{Bertini2022Bogoliubov}). In practice, one seeks a correction $\delta Q$ solving the cohomological equation $[H_{\rm nn},\delta Q]+[\hat H_{\rm nnn},Q]=0$, which admits a quasilocal solution $\delta Q$ of the same perturbative order as $\hat H_{\rm nnn}$, whenever the conserved charge $Q$ of $H_{\rm nn}$ is even under spin flip (i.e., for the charges $\hat Q_{2m-1}^{(1)}$ and $\hat Q_{2m}^{(2)}$, $m=1,2,\dots$). In that case the deformed charge $Q+\delta Q$ is conserved by the full Hamiltonian up to higher order, indeed  $[H_{\rm nn}+H_{\rm nnn},Q+\delta Q]=[H_{\rm nnn},\delta Q]$ and $[H_{\rm nnn},\delta Q]$ is second order in the perturbation. By contrast, for charges that are odd under spin flip (i.e., for the charges $\hat Q_{2m}^{(1)}$ and $\hat Q_{2m-1}^{(2)}$, $m=1,2,\dots$) the cohomological equation has no quasilocal solution: those charges are therefore \emph{fragile} and are spoiled on much shorter time scales. Importantly, the present weak-integrability breaking problem is subtler than in Ref.~\cite{Surace_2023} because the perturbation $\hat H_{\rm nnn}$ does not eliminate the role of resonant processes. The key question becomes how efficiently a conservation law regularizes such divergences, i.e., how robust they are under perturbations in the sense of Ref.~\cite{Burgarth2021KAM}. Finally, we remark that the fragility of the magnetization current (which is conserved in the unperturbed model) is expected to have direct consequences for spin transport: once that conservation law is destabilized, the perturbed model crosses over from ballistic to diffusive spin dynamics.

\paragraph{\textbf{Numerical simulations with MPS ---}}
Numerical simulations presented in Figs.~\ref{Fig:Magnetization},\ref{Fig:VarianceIntegrMag}, \ref{Fig:NoLogGrowth} and~\ref{Fig:String:Operator} are performed using MPS, which are a category of many-body quantum states characterized by finite correlations and little entanglement.
We provide here a few technical details on the simulations, which employ the Julia-base library ITensors~\cite{itensor, itensor-r0.3}.
Numerics are performed on lattices of $L=48$ sites, time-evolution is trotterized with time-step $\delta t = 0.01$ and the maximum allowed bond dimension is $\chi_{\rm max}= 512$, which is never saturated; during the sweeps on the setups, singular values smaller than $\epsilon_{\rm dis} = 10^{-12}$ are discarded.
Numerical errors are under control and convergence has been verified performing simulations with varying parameter values in the ranges $\delta t \in \{ 0.005,0.01,0.02 \}$ and $\epsilon_{\rm dis} \in \{10^{-8}, 10^{-10}, 10^{-12}\}$.

\begin{figure}[t]
    \centering
    \includegraphics[width=\columnwidth]{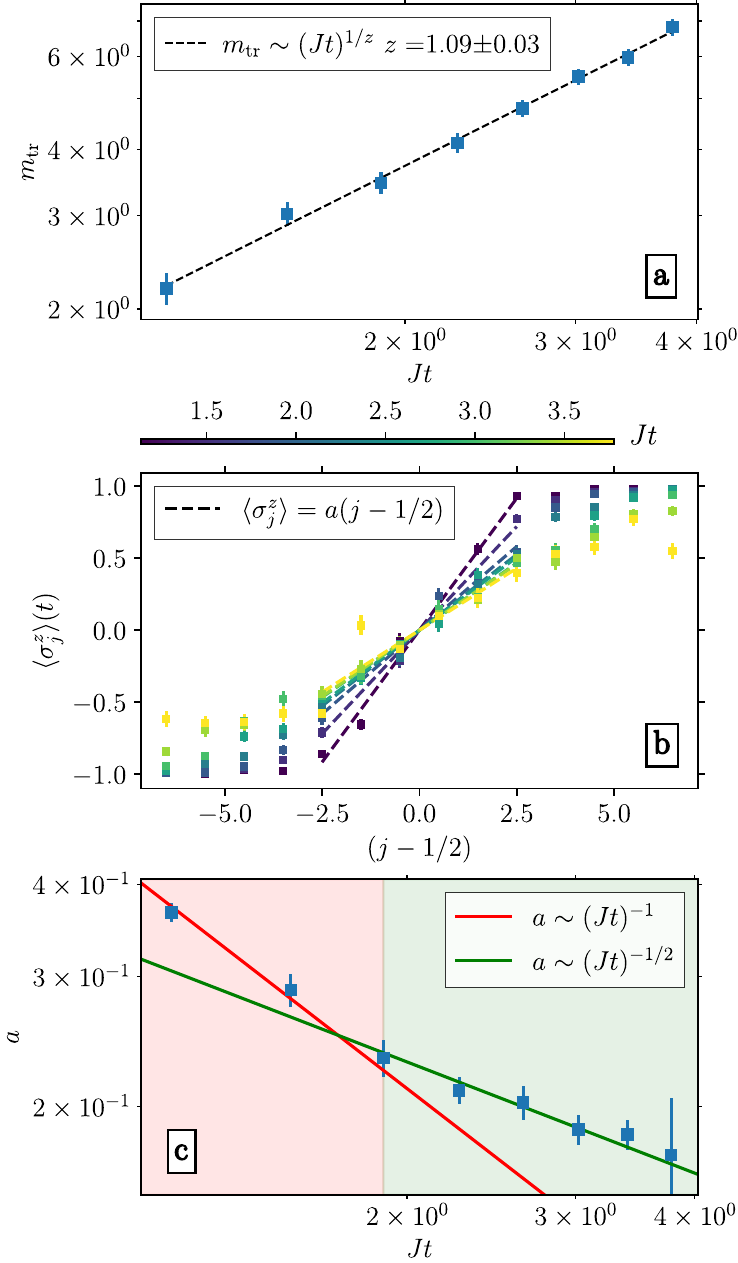}
    \caption{\textbf{Analysis of the magnetization profile.}
    \textbf{a}~The transferred magnetization grows as a power law $\sim t^{1/z}$ and the fitted value is compatible with ballistic transport.
    \textbf{b}~The experimental data for the magnetization profile are fitted around the central part of the chain with a straight line; here, the fit is performed considering $6$ points.
    \textbf{c}~Plot of the fitted slope as a function of time; the bi-logarithmic scale highlights a short-time ballistic scaling and the emergence of diffusion at late times.
    }
    \label{fig:SlopeAnalysis}
\end{figure}

\paragraph{\textbf{Quantitative analysis of the rescaling of the magnetization data ---}}
To assess which of the two rescalings proposed in Figs.~\ref{Fig:Magnetization}\textbf{e} and~\ref{Fig:Magnetization}\textbf{f} provides the better description for the late-time behaviour of the experimental $\langle \sigma^z_j \rangle (t)$, we compute the transferred magnetization, following a standard recipe:
\begin{equation}
    m_{\rm tr}(t) = - \langle \sum_{j} \text{sign} (j) \hat \sigma^z_j
    \rangle_t + \langle \sum_{j} \text{sign} (j) \hat \sigma^z_j \rangle_{t=0},
\end{equation}
where $\operatorname{sign}(j) = -1$ for $j \le 0$ and $1$ otherwise.
This quantity is plotted in Fig.~\ref{fig:SlopeAnalysis}\textbf{a} and is expected to grow as $t^{1/z}$; the fit performed on the experimental data yields $z=1.09 \pm 0.03$ on times in the interval $Jt \in [ 1.13, 3.78 ]$. This way of analyzing the data thus strongly suggests a ballistic behavior.

To show that this result is however not conclusive, we study the behavior of the magnetization profile near the junction. 
In both the ballistic and diffusive scenarios, the profile is smooth on the corresponding spatial scale and is therefore approximately linear in a small interval around the center of the chain. 
The two hypotheses, however, predict different time dependences for the central slope: in the ballistic case the slope scales as $\sim 1/t$, whereas in the diffusive case it scales as $\sim t^{-1/2}$. 

We extract the slope from the experimental data by fitting the data $\langle \sigma^z_j\rangle (t)$ in a symmetric region of $6$ sites around the junction at several times, as displayed in Fig.~\ref{fig:SlopeAnalysis}\textbf{b}.
Figure~\ref{fig:SlopeAnalysis}\textbf{c} shows the fitted slope as a function of time. The bi-logarithmnic scale highlights two different scalings, one $\sim t^{-1}$ at short times, and one $\sim t^{-1/2}$ at late times.
Whereas the initial ballistic behaviour depends on the number of sites employed to fit the slope, the diffusive behaviour appears to be more stable.
The slope of the magnetization profile, thus highlights the appearance of a diffusive behaviour that is consistent with the theoretical analyses for large spin chains.

\paragraph{\textbf{Experimental magnetization fluctuations ---}}

The data shown in Fig.~\ref{Fig:NoLogGrowth}\textbf{c}--\textbf{e} exhibit a
time-dependent increase that is qualitatively reproduced by our finite-size
simulations of the coherent dynamics generated by $\hat H_{\rm nn}+\hat H_{\rm nnn}$.
At later times, $t \gtrsim 5\,J^{-1}$ (corresponding to $\sim 730$~ns), the
experimental curves display an additional, abrupt upturn that is not captured by
the coherent simulations. We therefore refrain from assigning physical significance
to this late-time feature. In the figures we restrict the analysis to the time
window where experiment and coherent numerics are consistent, and we attribute the
unexplained late-time growth to accumulated experimental imperfections and other
systematic effects beyond our present control.

\paragraph{\textbf{The classical stochastic cellular automaton: the case $p=0$ ---}}

The $p=0$ cellular automaton is sufficiently simple that its dynamics can be solved
analytically, without resorting to numerical simulation. Initially, the right half
of the chain is fully occupied by quasiparticles. Their direction of motion is
fixed by the site parity: depending on $j\bmod 2$, a particle acts as a right- or a
left-mover. This parity dependence is apparent already in the first update steps
sketched in Fig.~\ref{Fig:Cellular:Sketch}.

As seen in Fig.~\ref{Fig:Authomaton}\textbf{a$_1$}, two regions persist outside the
light cone in which the density remains equal to its initial value, namely $q=0$ on
the left and $q=1$ on the right. Inside the light cone a nonequilibrium steady state
forms, consisting of an alternating pattern of occupied and empty sites. Its
coarse-grained density is therefore $1/2$, and it is composed solely of left-movers.
The particle current vanishes outside the light cone: the left region is empty,
while the right region contains right- and left-movers in equal proportion.

Both the density and the current thus admit a simple ballistic scaling form in terms
of the rescaled variable $j/t$:
\begin{subequations}
\begin{equation}
 q_j(t)=
 \begin{cases}
 0 & \text{for } j/t<-1,\\[1mm]
 1/2 & \text{for } -1< j/t < 1,\\[1mm]
 1 & \text{for } j/t>1,
 \end{cases}
\end{equation}
\begin{equation}
 \mathcal{J}_j(t)=
 \begin{cases}
 0 & \text{for } j/t<-1 \text{ or } j/t>1,\\[1mm]
 -1 & \text{for } -1< j/t < 1.
 \end{cases}
\end{equation}
\end{subequations}
Thus, the cellular automaton exhibits strictly ballistic
transport in the deterministic limit $p=0$.

\paragraph{\textbf{Deriving the telegrapher equation ---}}

To clarify the connection between the stochastic cellular automaton and the
telegrapher equation, it is convenient to consider a closely related model defined
directly in continuous space and time, namely the \emph{run-and-tumble process}%
~\cite{Angelani-15,bmrs-20,Jode-23}. In the dilute-scattering regime $p\ll 1$, this
process provides a natural continuum counterpart of the automaton discussed above.
Configurations are specified by the positions of two species of particles, right-
and left-movers. Each particle propagates ballistically with unit speed and flips
the sign of its velocity at a Poisson rate $\Gamma$, which plays a role analogous to
the backscattering probability $p$ in the discrete dynamics.

The hydrodynamics of the run-and-tumble process is well understood (and has also
appeared in the context of stochastic conformal field theory; see Ref.~\cite{bd-17}).
Denoting by $q_R(x,t)$ and $q_L(x,t)$ the coarse-grained densities of right- and
left-movers, it is convenient to introduce the total density and current, $q(x,t)=q_R(x,t)+q_L(x,t)$, $\mathcal J(x,t)=q_R(x,t)-q_L(x,t)$.
At the hydrodynamic level one obtains the coupled equations
\begin{equation}
\begin{cases}
\partial_t q(x,t)+\partial_x \mathcal J(x,t)=0,\\
\partial_t \mathcal J(x,t)+\partial_x q(x,t)=-2\Gamma\,\mathcal J(x,t),
\end{cases}
\end{equation}
where the first line is the continuity equation and the second describes the
relaxation of the current due to backscattering between the two velocity sectors.
Combining the two relations yields the telegrapher equation. Indeed, differentiating
the continuity equation with respect to time and using the second equation to
eliminate $\partial_t \mathcal J$ gives
Eq.~\eqref{Eq:Telegrapher}.

\bibliography{bibliography}

\textbf{Acknowledgements:}
We acknowledge fruitful discussions on the subject with F.~Ferro and G.~Morettini.
\textbf{Funding:}
This work is supported by
the ANR project LOQUST ANR-23-CE47-0006-02 and by the PEPR Dyn-1D ANR-23-
PETQ-0001 and the ANR-22-PETQ-0004 France 2030, project QuBitAF. This work is part of HQI (www.hqi.fr) initiative and is supported by France
2030 under the French National Research Agency grant number ANR-22-PNCQ-0002.
This work is also supported by the European Research Council (Advanced grant No. 101018511-ATARAXIA), 
and the Horizon Europe programme HORIZON-CL4- 2022-QUANTUM-02-SGA (project 101113690 (PASQuanS2.1)).
C.C.~acknowledges the support from Quantum Science and Technology-National Science and Technology Major Project (2024ZD0301700) and the start-up grant from IOP-CAS.

\textbf{Author contributions:}
C.C.,~G.B.~and G.E.~performed the experiment;
C.C.~also analyzed the data;
L.C.~conceived and performed the theoretical analysis based on the stochastic automaton;
A.M.~performed the numerical simulations.
T.L.,~A.B., M.F.~and L.M.~supervised the experimental and theoretical work.
All authors contributed to the analysis and interpretation of the experimental and theoretical data, and to the writing of the paper.
\textbf{Competing interests:} The authors declare that they have no competing interests.
\textbf{Data availability:} The experimental and numerical data presented in this paper are available upon reasonable request to the corresponding author. 

\end{document}